\documentclass[twocolumn,showpacs,preprintnumbers,amsmath,amssymb]{revtex4}
\usepackage{graphicx}% Include figure files
\usepackage{dcolumn}% Align table columns on decimal point
\usepackage{bm}% bold math

\usepackage{amssymb}
\usepackage{amsmath}

\begin{document}

%%\preprint{for Phys. Rev. D (\textbf{Ver. 04a})}

\title{Evolution of thermodynamic quantities on cosmological horizon in $\Lambda(t)$ model}
%%%\subtitle{}

\author{Nobuyoshi {\sc Komatsu}}  \altaffiliation{E-mail: komatsu@se.kanazawa-u.ac.jp} 
\affiliation{Department of Mechanical Systems Engineering, Kanazawa University, Kakuma-machi, Kanazawa, Ishikawa 920-1192, Japan}

%%\date{\today}

\begin{abstract}
The horizon of a flat Friedmann--Robertson--Walker (FRW) universe is considered to be dynamic when the Hubble parameter $H$ and the Hubble radius $r_{H}$ vary with time, unlike for de Sitter universes.
To clarify the thermodynamics on a dynamic horizon, the evolution of a dynamical Kodama--Hayward temperature and Bekenstein--Hawking entropy on the horizon of a flat FRW universe is examined in a $\Lambda(t)$ model similar to time-varying $\Lambda(t)$ cosmologies.
The $\Lambda(t)$ model includes both a power-law term proportional to $H^{\alpha}$ (where $\alpha$ is a free variable) and the equation of state parameter $w$, extending a previous analysis [Phys.\ Rev.\ D \textbf{100}, 123545 (2019)].
Using the present model, a matter-dominated universe ($w=0$) and a radiation-dominated universe ($w=1/3$) are examined, setting $\alpha <2$.
Both universes tend to approach de Sitter universes and satisfy the maximization of entropy in the last stage.
The evolution of several parameters (such as the Bekenstein--Hawking entropy) is similar for both $w=0$ and $w=1/3$, though the dynamical temperature $T_{H}$ is different.
In particular, $T_{H}$ is found to be constant when $w=1/3$ with $\alpha=1$, although $H$ and $r_{H}$ vary with time.
To discuss this case, the specific conditions required for constant $T_{H}$ are examined.
Applying the specific condition to the present model gives a cosmological model that can describe a universe at constant $T_{H}$, as if the dynamic horizon is in contact with a heat bath.
The relaxation processes for the universe are also discussed.

%%\keywords{First keyword \and Second keyword \and More}
%\PACS{98.80.-k, 98.80.Es, 95.30.Tg}
\end{abstract}
\maketitle

\section{Introduction} 
\label{Introduction}

To explain the accelerated expansion of the late Universe \cite{PERL1998_Riess1998,Planck2018,Hubble2017}, various cosmological models have been proposed  \cite{Weinberg1Roy1,Bamba1Nojiri1,HDE_review_Frusciante}, 
such as lambda cold dark matter ($\Lambda$CDM) models, time-varying $\Lambda (t)$ cosmology \cite{Freese-Sola2013,Nojiri2006-2017,Sola_2009-2018,Valent2015Sola2019,Sola_2017-2020}, bulk viscous cosmology \cite{Weinberg0,BarrowLima,BrevikNojiri,Meng3,Avelino2etc2020}, 
creation of CDM models \cite{Prigogine_1988-1989,Lima1992-1996,LimaOthers2001-2016}, and thermodynamic scenarios \cite{HDE,Easson,Cai,Basilakos2012-2016,Sheykhi1,Sheykhi2Karami,Koma4,Koma5,Koma678,Koma9,Koma10,Paul20212022,Nojiri2022etc}.
Most of the models imply that our Universe finally approaches a $\Lambda$-dominated universe, namely a de Sitter universe.
The de Sitter universe is in thermal equilibrium from the viewpoint of horizon thermodynamics \cite{GibbonsHawking1977}, which is closely related to black hole thermodynamics \cite{Bekenstein1,Hawking1,Hawking2}.

The thermodynamic scenario and thermodynamics of the universe have been extensively examined \cite{Jacob1995,Padma2010,Verlinde1,Padmanabhan2004,ShuGong2011,Padma2012AB,Cai2012-Tu2013,Tu2013-2015,Sheykhia2018,Neto2018a,Krishna2023,Easther1,Barrow3,Davies11_Davis0100,Gong00_01,Egan1,deSitter,Pavon2013Mimoso2013,Krishna20172019,Bamba2018Pavon2019,Saridakis20192021,Koma11,Koma12,Koma14,Koma15,Koma16,Koma17,Koma18}, 
especially based on the holographic principle \cite{Hooft-Bousso}.
In those works, the Gibbons--Hawking temperature \cite{GibbonsHawking1977} is widely used as an approximate temperature on the cosmological horizon.
The Gibbons--Hawking temperature is constant during evolution of de Sitter universes, in which the Hubble radius and the Hubble parameter are also constant.
In contrast, these three quantities vary with time in the late Universe \cite{Hubble2017}.
In this sense, the horizon of the de Sitter universe is static, whereas horizons of other universes (including our Universe) are generally considered to be dynamic.

In fact, a dynamical temperature (called the Kodama--Hayward temperature) has been proposed to describe the temperature on dynamic horizons of black holes and universes \cite{Dynamical-T-1998,Dynamical-T-2008,Dynamical-T-20072014}.
The dynamical temperature on the cosmological horizon \cite{Dynamical-T-20072014} is considered to be an extended Gibbons--Hawking temperature and has been examined from various viewpoints \cite{Tu2018,Tu2019,ApparentHorizon2022,Mathew2023}.
The dynamical temperature should be suitable for discussing the thermodynamics on a dynamic horizon.
However, the evolution of the dynamical temperature has not yet been sufficiently studied in cosmological models.

We therefore examine the evolution of the dynamical temperature $T_{H}$ on the horizon of a flat Friedmann--Robertson--Walker (FRW) universe.
For cosmological models, we consider a $\Lambda(t)$ model \cite{Koma11,Koma12,Koma14,Koma15}, similar to a time-varying $\Lambda(t)$ cosmology, which is a commonly used model \cite{Koma16}.
The $\Lambda(t)$ model includes a power-law term proportional to $H^{\alpha}$, where $H$ is the Hubble parameter and $\alpha$ is a free parameter \cite{Koma11}.
Although this model has been used for a matter-dominated universe ($w=0$) \cite{Koma14,Koma15,Koma16}, a radiation-dominated universe ($w=1/3$) has not yet been examined, where $w$ represents the equation of state parameter.
Naturally, a dynamical temperature was not discussed in the earlier works.
Therefore, it is worth examining the evolution of $T_{H}$ in matter-dominated and radiation-dominated universes in the $\Lambda(t)$ model.
In addition, we recently found that a universe with constant $T_{H}$ is related to a radiation-dominated universe in a $\Lambda(t)$ model.
The constant $T_{H}$ universe should extend the concept of horizons at constant temperature and may provide new insights for the discussion of horizon thermodynamics.

In this context, we examine the horizon thermodynamics of matter-dominated and radiation-dominated universes in the $\Lambda(t)$ model by observing the dynamical temperature $T_{H}$ and the Bekenstein--Hawking entropy.
The $\Lambda(t)$ model used here includes both a power-law term and the equation of state parameter, extending previous analyses \cite{Koma11,Koma12,Koma14,Koma15,Koma16}.
In addition, we study cosmological models that can describe a universe at constant $T_{H}$.

The remainder of the present article is organized as follows.
In Sec.\ \ref{Thermodynamics on the horizon}, horizon thermodynamics is reviewed.
The Bekenstein--Hawking entropy and the dynamical temperature $T_{H}$ on the cosmological horizon are introduced.
In Sec.\ \ref{Lambda(t) model with a power-law term}, we introduce a $\Lambda(t)$ model that includes both a power-law term and the equation of state parameter.
Using the present model, we examine the evolution of the Bekenstein--Hawking entropy and the dynamical temperature $T_{H}$.
In Sec.\ \ref{Constant T}, we study the specific conditions required for constant $T_{H}$ on dynamic horizons.
Based on the specific conditions and the present model, we formulate a cosmological model that can describe a universe at constant $T_{H}$.
We also discuss the properties of the universe in the formulated model.
Finally, in Sec.\ \ref{Conclusions}, the conclusions of the study are presented.

In this paper, a flat FRW universe is considered and, therefore, the Hubble horizon is equivalent to an apparent horizon.
An expanding universe is assumed as well.
Inflation of the early universe and density perturbations related to structure formations are not discussed.

\section{Horizon thermodynamics} 
\label{Thermodynamics on the horizon}

The horizon of a universe is assumed to have an associated entropy and an approximate temperature \cite{Easson}, based on the holographic principle \cite{Hooft-Bousso}.
The entropy and the temperature are introduced in this section.

We select the Bekenstein--Hawking entropy as the associated entropy \cite{Bekenstein1,Hawking1,Hawking2}.
In general, the cosmological horizon is examined by replacing the event horizon of a black hole by the cosmological horizon \cite{Koma17,Koma18}. 
This replacement method has been widely accepted \cite{Jacob1995,Padma2010,Verlinde1,HDE,Padma2012AB,Cai2012-Tu2013,Tu2013-2015,Sheykhia2018,Neto2018a,Krishna2023,Padmanabhan2004,ShuGong2011}
and we use it here.

Based on the form of the Bekenstein--Hawking entropy, the entropy $S_{\rm{BH}}$ on the Hubble horizon is written as 
\begin{equation}
S_{\rm{BH}}  = \frac{ k_{B} c^3 }{  \hbar G }  \frac{A_{H}}{4}   ,
\label{eq:SBH}
\end{equation}
where $k_{B}$, $c$, $G$, and $\hbar$ are the Boltzmann constant, the speed of light, the gravitational constant, and the reduced Planck constant, respectively.
The reduced Planck constant is defined by $\hbar \equiv h/(2 \pi)$, where $h$ is the Planck constant \cite{Koma11,Koma12}.
$A_{H}$ is the surface area of the sphere with a Hubble horizon (radius) $r_{H}$ given by
\begin{equation}
     r_{H} = \frac{c}{H}   , 
\label{eq:rH}
\end{equation}
where the Hubble parameter $H$ is defined by 
\begin{equation}
   H \equiv   \frac{ da/dt }{a(t)} =   \frac{ \dot{a}(t) } {a(t)}  , 
\label{eq:Hubble}
\end{equation}
and $a(t)$ is the scale factor at time $t$ \cite{Koma11}.
Substituting $A_{H}=4 \pi r_{H}^2 $ into Eq.\ (\ref{eq:SBH}) and applying Eq.\ (\ref{eq:rH}) yields
\begin{equation}
S_{\rm{BH}}  = \frac{ k_{B} c^3 }{  \hbar G }   \frac{A_{H}}{4}       
                  =  \left ( \frac{ \pi k_{B} c^5 }{ \hbar G } \right )  \frac{1}{H^2}  
                  =    \frac{K}{H^2}    , 
\label{eq:SBH2}      
\end{equation}
where $K$ is a positive constant given by
\begin{equation}
  K =  \frac{  \pi  k_{B}  c^5 }{ \hbar G } . 
\label{eq:K-def}
\end{equation}
The normalized $S_{\rm{BH}}$ is written as \cite{Koma15}
\begin{equation}
\frac{S_{\rm{BH}}}{ S_{\rm{BH},0} } =  \left ( \frac{ H }{ H_{0} } \right )^{-2}  , 
\label{eq:SBHSBH0_0}      
\end{equation}
where the subscript $0$ represents the present time $t_{0}$.

When a de Sitter universe is considered, $r_{H}$ and $S_{\rm{BH}}$ are constant during the evolution of the universe because $H$ is constant.
In this sense, the horizon of the de Sitter universe is considered to be static.
Note that the scale factor for the de Sitter universe varies with time \cite{Koma17}: 
\begin{equation}  
      \frac{a}{a_{0}}  =    \exp[ H ( t - t_{0} ) ]     ,
\label{eq:aa0_deSitter}
\end{equation}
where $a_{0}$ represents the scale factor at the present time.

Next, we introduce an approximate temperature on the Hubble horizon.
Before introducing the dynamical temperature, we will review the Gibbons--Hawking temperature.
The Gibbons--Hawking temperature $T_{\rm{GH}}$ is given by \cite{GibbonsHawking1977} 
\begin{equation}
T_{\rm{GH}}  = \frac{ \hbar H}{   2 \pi  k_{B}  }   .
\label{eq:T_H1}
\end{equation}
This equation indicates that $T_{\rm{GH}}$ is proportional to $H$ and is constant during the evolution of de Sitter universes.
In fact, $T_{\rm{GH}}$ is obtained from field theory in the de Sitter space \cite{GibbonsHawking1977}.
However, most universes are not pure de Sitter universes in that their horizons are dynamic.
A similar dynamic horizon for black holes has been examined in the works of Hayward \cite{Dynamical-T-1998} and Hayward \textit{et al}. \cite{Dynamical-T-2008}.
Hayward suggested a dynamical temperature on a black hole horizon and clarified the relationship between the surface gravity and the temperature on a dynamic apparent horizon for the Kodama observer \cite{Dynamical-T-1998}.
(The Kodama--Hayward temperature was discussed in, e.g., the recent work of Muhsinath \textit{et al}. \cite{Mathew2023}.)

Based on the works of Hayward \textit{et al.}, a dynamical temperature on the cosmological horizon of an FRW universe has been proposed \cite{Dynamical-T-20072014} and examined from various viewpoints \cite{Tu2018,Tu2019,ApparentHorizon2022,Mathew2023}.
When a flat universe is considered, the apparent horizon is equivalent to the Hubble horizon.
Consequently, the dynamical temperature $T_{H}$ for a flat FRW universe can be written as \cite{Tu2018,Tu2019}
\begin{equation}
 T_{H} = \frac{ \hbar H}{   2 \pi  k_{B}  }  \left ( 1 + \frac{ \dot{H} }{ 2 H^{2} }\right )  ,
\label{eq:T_H_mod}
\end{equation}
where $H>0$ is used for an expanding universe.
For de Sitter universes, $T_{H}$ reduces to $T_{\rm{GH}}$.
That is, $T_{H}$ is considered to be an extended version of $T_{\rm{GH}}$.
For details of $T_{H}$, see, e.g., the works of Tu \textit{et al}. \cite{Tu2018,Tu2019}.

In this study, based on Eq.\ (\ref{eq:T_H_mod}), we consider the normalized temperature:
\begin{equation}
\frac{T_{H}}{ T_{\rm{GH},0} } =  \frac{H}{H_{0}}  \left ( 1 + \frac{ \dot{H} }{ 2 H^{2} }\right )   ,
\label{eq:T_H_mod_0}      
\end{equation}
where $T_{\rm{GH},0}$ is the Gibbons--Hawking temperature at the present time, given by $T_{\rm{GH},0} = \frac{ \hbar H_{0} }{   2 \pi  k_{B}  } $.
In the next section, the normalized entropy and the normalized temperature are examined, using a $\Lambda(t)$ model.

We note that various black hole entropies have been proposed by extending the Bekenstein--Hawking entropy \cite{Das2008Radicella2010,MeissnerGhosh,Tsallis2012,Czinner1,Czinner2,Barrow2020}.
The thermodynamic consistency of non-Gaussian black-hole entropies has been examined in Ref.\ \cite{Nojiri2021}.
Those entropies have been applied to dynamic horizons of universes, see, e.g., Refs.\ \cite{Sheykhi2Karami,Koma5,Koma10,Paul20212022,Nojiri2022etc,Koma11}.
While it is worthwhile studying the thermodynamic relations between the dynamical temperature and the entropy on the cosmological horizon,
the thermodynamic relation is not discussed here and 
the present study focuses on and examines evolution of thermodynamic quantities.

\section{$\Lambda(t)$ model with a power-law term} 
\label{Lambda(t) model with a power-law term}

We review the $\Lambda(t)$ model with a power-law term and study the evolution of the Bekenstein--Hawking entropy $S_{\rm{BH}}$ and the dynamical temperature $T_{H}$.
In Sec.\ \ref{General formulation}, the $\Lambda(t)$ model is introduced.
In Sec.\ \ref{Deceleration parameter for the present model}, background evolution of the universe for the present model are discussed.
The evolution of the entropy and the temperature is examined in Secs.\ \ref{Entropy for the present model} and \ref{Temperature for the present model}, respectively.
We consider a flat FRW universe and assume an expanding universe.

\subsection{Cosmological equations} 
\label{General formulation}

Based on previous works \cite{Koma14,Koma15,Koma16,Koma17}, a $\Lambda(t)$ model that includes both a power-law term and the equation of state parameter is introduced, using a general formulation of the cosmological equations.
The general Friedmann equation for the $\Lambda(t)$ model is given as 
\begin{equation}
 H(t)^2      =  \frac{ 8\pi G }{ 3 } \rho (t)    + f_{\Lambda}(t)            ,                                                 
\label{eq:General_FRW01_f_0} 
\end{equation} 
and the general acceleration equation is 
\begin{align}
  \frac{ \ddot{a}(t) }{ a(t) }      &=  -  \frac{ 4\pi G }{ 3 }  ( 1+  3w ) \rho (t)     +   f_{\Lambda}(t)                ,  
\label{eq:General_FRW02_f_0}
\end{align}
where $w$ represents the equation of state parameter for a generic component of matter, $w = p(t) / (\rho(t)  c^2)$.
Also, $\rho(t)$ and $p(t)$ are the mass density and pressure of cosmological fluids, respectively \cite{Koma14,Koma16,Koma17}.
For a matter-dominated universe, a radiation-dominated universe, and a $\Lambda$-dominated universe, $w$ is $0$, $1/3$, and $-1$, respectively.
In this paper, $w=0$ and $w=1/3$ are considered.
An extra driving term $f_{\Lambda}(t)$ is phenomenologically assumed.
Combining Eq.\ (\ref{eq:General_FRW01_f_0}) with Eq.\ (\ref{eq:General_FRW02_f_0}) yields \cite{Koma14}
\begin{equation}
      \dot{H} = - \frac{3}{2} (1+w) H^{2}  +  \frac{3}{2} (1+w)    f_{\Lambda}(t)     .  
\label{eq:Back_f}
\end{equation}

Using the above equation, we have phenomenologically formulated a $\Lambda(t)$ model that includes a power-law term based on Padmanabhan's holographic equipartition law \cite{Koma11,Koma14,Koma15,Koma16}.
The power-law term has been investigated in previous works \cite{Koma14,Koma15,Koma16}.
According to these works, we use the following power-law term:
\begin{equation}
        f_{\Lambda}(t)   =   \Psi_{\alpha} H_{0}^{2} \left (  \frac{H}{H_{0}} \right )^{\alpha}  , 
\label{eq:fLpower}
\end{equation}
where $\alpha$ and $\Psi_{\alpha}$ are dimensionless constants whose values are real numbers \cite{Koma11}.
Also, $\alpha$ and $\Psi_{\alpha}$ are independent free parameters, and $\alpha < 2$ and $0 \leq \Psi_{\alpha} \leq 1 $ are considered.
That is, $\Psi_{\alpha}$ is a kind of density parameter for the effective dark energy.
For the derivation of the power-law term, see, e.g., Ref.\ \cite{Koma11}.
A similar power series for $H$ in $\Lambda(t)$ models was examined in Ref.\ \cite{Valent2015Sola2019}.

Substituting Eq.\ (\ref{eq:fLpower}) into Eq.\ (\ref{eq:Back_f}) yields 
\begin{align}
    \dot{H} &= - \frac{3}{2} (1+w)  H^{2}  +  \frac{3}{2}   (1+w)  \Psi_{\alpha} H_{0}^{2} \left (  \frac{H}{H_{0}} \right )^{\alpha}             \notag \\
                &= - \frac{3 (1+w) }{2} H^{2}  \left (  1 -   \Psi_{\alpha} \left (  \frac{H}{H_{0}} \right )^{\alpha -2} \right )        .
\label{eq:Back_power_1}
\end{align}
This equation is satisfied for all $\alpha$ \cite{Koma14}.
The solutions can be categorized according to whether or not $\alpha =2$.
The solution for $\alpha = 2$ is written as \cite{Koma14}
\begin{equation}
     \frac{H}{H_{0}}  =     \left ( \frac{a}{a_{0}} \right )^{ - \frac{3 (1+w) (1- \Psi_{\alpha}) }{2}  }    ,
\label{eq:Sol_HH0_aa0_H2_2w}
\end{equation}
and the solution for $\alpha \neq 2$ is written as 
\begin{equation}  
    \left ( \frac{H}{H_{0}} \right )^{2-\alpha}  =   (1- \Psi_{\alpha})   \left ( \frac{a}{a_{0}} \right )^{ - \frac{3 (1+w) (2-\alpha)}{2}  }  + \Psi_{\alpha}      .
\label{eq:Sol_HH0_power_00}
\end{equation}
The solution method is summarized in Ref.\ \cite{Koma14}.
This model has been studied for $w=0$ in Refs.\ \cite{Koma14,Koma15}
and it was found that when $w=0$, $\alpha <2$ leads to an initially decelerating and then accelerating universe (hereafter a `decelerating and accelerating universe').
Also, when $w=0$, the universe for $\alpha < 2$ satisfies the maximization of entropy in the last stage \cite{Koma14,Koma15}.
Therefore, $\alpha < 2$ is considered in this study.

Using the normalized scale factor $\tilde{a}$, the solution for $\alpha \neq 2$ given by Eq.\ (\ref{eq:Sol_HH0_power_00}) is written as
\begin{equation}  
    \left ( \frac{H}{H_{0}} \right )^{2-\alpha}  =   (1- \Psi_{\alpha})   \tilde{a}^{- \gamma}   + \Psi_{\alpha}      ,
\label{eq:Sol_HH0_power}
\end{equation}
where $\tilde{a}$ and the parameter $\gamma$ are given by 
\begin{equation}
\tilde{a} = \frac{a}{a_{0}} \quad \textrm{and} \quad \gamma = \frac{3 (1+w) (2-\alpha)}{2} ,.
\label{aa0gama}
\end{equation}
A coefficient $(1+w)$ is included in $\gamma$.
In this paper, $\alpha <2$, $w=0$, and $w=1/3$ are considered.
Therefore, $2-\alpha$, $1+w$, and $\gamma$ are positive.

We note that $\Lambda$CDM models are obtained from Eq.\ (\ref{eq:Sol_HH0_power}), neglecting the influence of radiation.
Substituting $\alpha =0$ and $w =0$ into Eq.\ (\ref{aa0gama}) yields $\gamma =3$.
In addition, substituting $\alpha =0$ and $\gamma =3$ into Eq.\ (\ref{eq:Sol_HH0_power}) and replacing $\Psi_{\alpha}$ by $\Omega_{\Lambda}$ yields \cite{Koma14}
\begin{equation}
 \left (  \frac{H}{H_{0}} \right )^{2}  =   (1- \Omega_{\Lambda} )   \tilde{a}^{ - 3}  + \Omega_{\Lambda}    ,
\label{eq:Sol_H_LCDM}
\end{equation}
where $\Omega_{\Lambda}$ is the density parameter for $\Lambda$ and is given by $\Lambda /( 3 H_{0}^{2} ) $. 
The above equation corresponds to the $\Lambda$CDM model in a flat FRW universe, where the influence of radiation is neglected.

\subsection{Deceleration parameter $q$} 
\label{Deceleration parameter for the present model}

In this subsection, we examine the background evolution of the universe for the present model.
To this end, we observe the evolution of the Hubble parameter and a deceleration parameter
$q$, defined by
\begin{equation}
q \equiv  - \left ( \frac{\ddot{a} } {a H^{2}} \right )  , 
\label{eq:q_def0}
\end{equation}
where a positive or negative $q$ represents deceleration or acceleration, respectively \cite{Koma14}. 
Substituting $\ddot{a}/a = \dot{H} + H^{2}$ into Eq.\ (\ref{eq:q_def0}) and substituting Eq.\ (\ref{eq:Back_power_1}) into the resultant equation yields
\begin{align}
q &=   -  \frac{ \dot{H} } {H^{2}}   -1    =  \frac{3}{2} (1+w)   \left (  1 -   \Psi_{\alpha} \left (  \frac{H}{H_{0}} \right )^{\alpha -2} \right )             -1     .
\label{eq:q_power0}
\end{align}
Substituting Eq.\ (\ref{eq:Sol_HH0_power}) into the above equation yields
\begin{align}
q &=   \frac{3}{2} (1+w)   \left (  1 -   \frac{\Psi_{\alpha}}{ (1- \Psi_{\alpha})   \tilde{a}^{- \gamma}   + \Psi_{\alpha} } \right )                           -1     \notag \\
   &=      \frac{ \frac{3}{2} (1+w)  (1- \Psi_{\alpha})   \tilde{a}^{- \gamma} }{ (1- \Psi_{\alpha})   \tilde{a}^{- \gamma}   + \Psi_{\alpha} }              -1    ,
\label{eq:q_power02}
\end{align}
where $\gamma = \frac{3 (1+w) (2-\alpha)}{2}$, as given by Eq.\ (\ref{aa0gama}).
Equation\ (\ref{eq:q_power02}) includes $\gamma$ and a coefficient $(1+w)$.

\begin{figure} [t] 
\begin{minipage}{0.495\textwidth}
\begin{center}
\scalebox{0.33}{\includegraphics{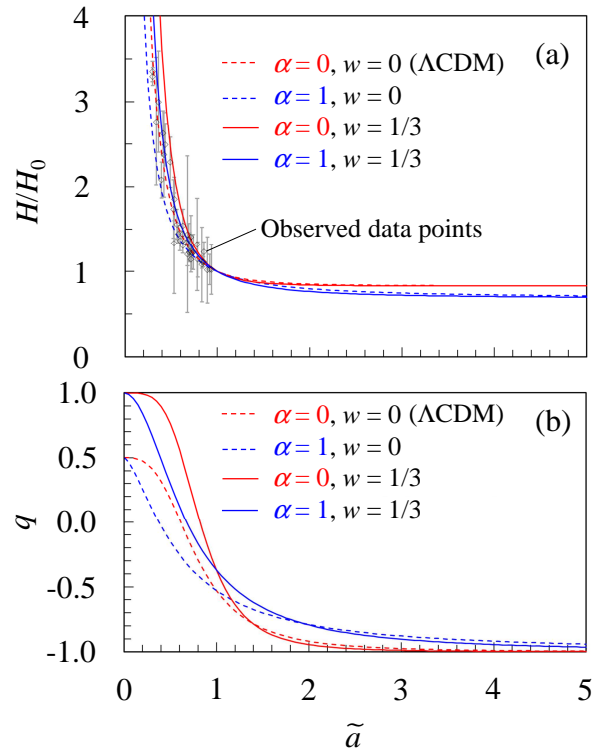}}
\end{center}
\end{minipage}
\caption{Evolution of the universe for the present model for $\Psi_{\alpha} =0.685$. 
(a) Normalized Hubble parameter $H/H_{0}$.
(b) Deceleration parameter $q$.
The dashed and solid lines represent $w=0$ and $w=1/3$, respectively.
The red and blue lines represent $\alpha =0$ and $\alpha =1$, respectively.
In (a), the open diamonds with error bars are observed data points taken from Ref.\ \cite{Hubble2017}. 
To normalize the data points, $H_{0}$ is set to $67.4$ km/s/Mpc from Ref.\ \cite{Planck2018}. 
Similar evolution for $w=0$ has been examined in Refs.\ \cite{Koma14,Koma15,Koma16,Koma17}.
 }
\label{Fig-H-a}
\end{figure}

Figure \ref{Fig-H-a} illustrates the evolution of the Hubble parameter and the deceleration parameter.
The dashed and solid lines represent $w=0$ and $w=1/3$, respectively.
To examine typical results, $\alpha$ is set to $0$ and $1$.
In addition, $\Psi_{\alpha}$ is set to $0.685$, which is equivalent to $\Omega_{\Lambda}$ for the $\Lambda$CDM model from the Planck 2018 results \cite{Planck2018}.
That is, the plots for [$\alpha =0$, $w=0$] are equivalent to those for the $\Lambda$CDM model.
The normalized scale factor $\tilde{a}$ increases with time because an expanding universe is considered.
Similar evolution for $w=0$ has been examined in Refs.\ \cite{Koma14,Koma15,Koma16,Koma17}.

As shown in Fig.\ \ref{Fig-H-a}(a), $H/H_{0}$ decreases with $\tilde{a}$ and gradually approaches a positive value that depends on $\alpha$ and $\Psi_{\alpha}$ but not on $w$.
The positive value is given by $H/H_{0} = \Psi_{\alpha}^{1/(2-\alpha)}$, which is obtained by applying $ \tilde{a} \rightarrow \infty$ to Eq.\ (\ref{eq:Sol_HH0_power}) with $\alpha < 2$ \cite{Koma17}.
Before approaching the positive value, $H/H_{0}$ for $w=1/3$ is quantitatively different from that for $w=0$.
However, the evolution of $H/H_{0}$ for $w=0$ and $w=1/3$ is similar.
We note that $H/H_{0}$ is equivalent to the normalized Gibbons--Hawking temperature $T_{\rm{GH}} = \hbar H/(2 \pi  k_{B})$, as given by Eq.\ (\ref{eq:T_H1}), because $T_{\rm{GH}}$ is proportional to $H$.

Next, we observe the evolution of the deceleration parameter $q$.
As shown in Fig.\ \ref{Fig-H-a}(b), $q$ decreases with $\tilde{a}$ and gradually approaches $-1$, although it is positive in the early stage.
Also, $q$ is negative at $\tilde{a}=1$, namely at the present time.
This result indicates a decelerating and accelerating universe, as examined in Refs.\ \cite{Koma14,Koma15}.
In addition, $q$ for $w=1/3$ is quantitatively different from that for $w=0$, but the evolution of $q$ for $w=1/3$ is similar to that for $w=0$.

The deceleration parameter $q$ depends on $w$, $\alpha$, $\Psi_{\alpha}$, and $H/H_{0}$, as shown in Eq.\ (\ref{eq:q_power0}).
Therefore, we discuss an accelerating universe using the $(\alpha, \Psi_{\alpha})$ plane.
The boundary required for $q = 0$ can be calculated from Eq.\ (\ref{eq:q_power0}).
(The boundary of $q= 0$ for $w=0$ was discussed in Ref.\ \cite{Koma14}.)
Substituting $q = 0$ into Eq.\ (\ref{eq:q_power0}) yields 
\begin{align}
   1 -   \Psi_{\alpha} \left (  \frac{H}{H_{0}} \right )^{\alpha -2}    &=  \frac{2}{3 (1+w) }      .
\label{q0_1}
\end{align}
Solving this with respect to $\Psi_{\alpha}$ yields 
\begin{align}
    \Psi_{\alpha} &=  \left ( 1- \frac{2}{3 (1+w) }   \right)  \left (  \frac{H}{H_{0}} \right )^{2- \alpha}   =  \frac{1+3w}{3 (1+w) }   \left (  \frac{H}{H_{0}} \right )^{2- \alpha}   .
\label{q0_2}
\end{align}
This equation is satisfied for all $\alpha$.
When $\alpha = 2$, from Eq.\ (\ref{q0_2}), the boundary (point) is given by 
\begin{align}
 \Psi_{\alpha} = \frac{1+3w}{3 (1+w)}     \quad   (\textrm{for}  \quad  \alpha = 2).
\label{q0_2_alpha2}
\end{align}
Also, substituting Eq.\ (\ref{eq:Sol_HH0_power}) into Eq.\ (\ref{q0_2}) yields 
\begin{align}
    \Psi_{\alpha} &=   \frac{1+3w}{3 (1+w) }   \left [    (1- \Psi_{\alpha})   \tilde{a}^{- \gamma} + \Psi_{\alpha}  \right ]   ,
\label{q0_3}
\end{align}
and solving Eq.\ (\ref{q0_3}) with respect to $\Psi_{\alpha}$ yields the following boundary of $q = 0$:
\begin{align}  
   \Psi_{\alpha} =  \frac{  (1+3w) \tilde{a}^{- \gamma}      }{  2  +  (1+3w) \tilde{a}^{- \gamma}      }  ,
\label{q0_psi}
\end{align}
where $\gamma = \frac{3 (1+w) (2-\alpha)}{2}$ from Eq.\ (\ref{aa0gama}).
The above equation includes not only $\gamma$ but also a coefficient $(1+3w)$.
In this study, $1+3w$ is positive because $w=0$ and $w=1/3$ are considered.

\begin{figure} [t] 
\begin{minipage}{0.495\textwidth}
\begin{center}
\scalebox{0.34}{\includegraphics{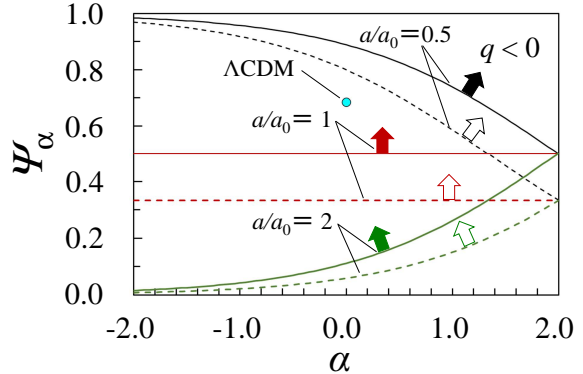}}
\end{center}
\end{minipage}
\caption{Boundary of $q= 0$ in the $(\alpha, \Psi_{\alpha})$ plane for various values of $a/a_{0}$. 
The dashed and solid lines represent $w=0$ and $w=1/3$, respectively.
The arrow attached to each boundary indicates an accelerating-universe-side region that satisfies $q< 0$.
The open circle represents $(\alpha, \Psi_{\alpha}) = (0, 0.685)$ for the $\Lambda$CDM model. }
\label{Fig-q_plane_power}
\end{figure}

Using Eq.\ (\ref{q0_psi}), the boundary of $q = 0$ can be plotted in the $(\alpha, \Psi_{\alpha})$ plane.
In Fig.\ \ref{Fig-q_plane_power}, $a/a_{0}$ is set to $0.5$, $1$, and $2$, to examine typical boundaries.
In an expanding universe, $a/a_{0}$ increases with time.
The arrow attached to each boundary indicates an accelerating-universe-side region that satisfies $q< 0$.
The upper side of each boundary corresponds to this region.
The dashed and solid lines represent $w=0$ and $w=1/3$, respectively.
Similar boundaries for $w=0$ have been examined in Ref.\ \cite{Koma14}.
In this figure, to avoid confusion, $a/a_{0}$ is used for the normalized scale factor, instead of $\tilde{a}$, because the symbol $\tilde{a}$ is similar to the symbol $\alpha$ on the horizontal axis.

As shown in Fig.\ \ref{Fig-q_plane_power}, the accelerating-universe-side region for both $w=0$ and $w=1/3$ varies with $a/a_{0}$.
The region for $w=0$ is similar to that for $w=1/3$. 
For example, in both cases, the boundaries for $a/a_{0}=0.5$ imply that a large-$\alpha$ and large-$\Psi_{\alpha}$ region tends to occur on the accelerating universe side.
In contrast, the boundaries for $a/a_{0}=2$ imply that a small-$\alpha$ and large-$\Psi_{\alpha}$ region tends to occur on the accelerating universe side.
In both cases, a decelerating and accelerating universe is further expected with increasing $a/a_{0}$.
The results are consistent with those in Ref.\ \cite{Koma14}.

Of course, the boundaries for $w=1/3$ are quantitatively different from those for $w=0$.
That is, the boundaries for $w=1/3$ are located higher than those for $w=0$.
To examine this difference, we observe the two boundaries for $a/a_{0}=1$, represented by the two horizontal lines in Fig.\ \ref{Fig-q_plane_power}.
When $a/a_{0}=1$, the boundary is given by $ \Psi_{\alpha} = \frac{1+3w}{3 (1+w)}$, which is obtained by applying $\tilde{a} =a/a_{0} =1$ to Eq.\ (\ref{q0_psi}).
The obtained boundary depends on $w$ and is equivalent to Eq.\ (\ref{q0_2_alpha2}).

\subsection{Entropy $S_{\rm{BH}}$ on the horizon} 
\label{Entropy for the present model}

Ordinary, isolated macroscopic systems spontaneously evolve to equilibrium states of maximum entropy consistent with their constraints \cite{Callen}.
Previous works imply that certain types of universe behave as ordinary macroscopic systems \cite{Pavon2013Mimoso2013,deSitter,Krishna20172019,Bamba2018Pavon2019,Saridakis20192021,Koma14,Koma15}.
In other words, the entropy on a cosmological horizon does not decrease, i.e., $\dot{S}_{\rm{BH}} \geq 0$.
Also, the entropy approaches a certain maximum value in the last stage, that is, the maximization of entropy, $\ddot{S}_{\rm{BH}} < 0$, should be satisfied.

In this subsection, we examine the entropy $S_{\rm{BH}}$ on the horizon for the present model for both $w=0$ and $w=1/3$.
From Eq.\ (\ref{eq:SBHSBH0_0}), the normalized $S_{\rm{BH}}$ is written as 
\begin{equation}
\frac{S_{\rm{BH}}}{ S_{\rm{BH},0} } =  \left ( \frac{ H }{ H_{0} } \right )^{-2}  . 
\label{eq:SBHSBH0_0a}      
\end{equation}
Substituting Eq.\ (\ref{eq:Sol_HH0_power}) into Eq.\ (\ref{eq:SBHSBH0_0a}) yields 
\begin{align}  
\frac{S_{\rm{BH}}}{ S_{\rm{BH},0} } &=    \left [     (1- \Psi_{\alpha})  \tilde{a}^{ - \gamma  }  + \Psi_{\alpha}     \right ]^{\frac{2}{\alpha-2}}      ,
\label{eq:SBH_power}      
\end{align}  
where $\gamma$ is $\frac{3 (1+w) (2-\alpha)}{2}$ given by Eq.\ (\ref{aa0gama}), which includes a coefficient $(1+w)$.
The case for $w=0$ has been discussed in previous works \cite{Koma14,Koma15}.

The calculations of $\dot{S}_{\rm{BH}}$ and $\ddot{S}_{\rm{BH}}$ are summarized in Appendix \ref{SL}, extending previous analyses \cite{Koma14,Koma15}.
Using Eq.\ (\ref{eq:dSBH_2_3_power_Complicate}) and $S_{\rm{BH},0}= K/H_{0}^{2}$, we obtain the normalized $\dot{S}_{\rm{BH}}$, which is given by
\begin{align}  
\frac{\dot{S}_{\rm{BH}}}{ S_{\rm{BH},0} H_{0}}  &=  \frac{  3 (1+w) (1- \Psi_{\alpha})   \tilde{a}^{- \gamma}     }{   \left [ (1- \Psi_{\alpha})   \tilde{a}^{- \gamma}   + \Psi_{\alpha}  \right ]^{ \frac{3- \alpha}{2- \alpha} }  } .
\label{eq:dSBH_2_3_power_Complicate00}      
\end{align}  
This equation indicates that $\dot{S}_{\rm{BH}} \geq 0$ is satisfied because\ $w= 0$, $w=1/3$, and $0 \le \Psi_{\alpha} \le 1$ are considered.
In addition, from Eq.\ (\ref{eq:d2Sdt2_power_1_Complicate}), the normalized $\ddot{S}_{\rm{BH}}$ is written as
\begin{align} 
   \frac{ \ddot{S}_{\rm{BH}}  }{S_{\rm{BH},0} H_{0}^{2} } &=  \frac{9}{2}   (1+w)^{2}  (1- \Psi_{\alpha})    \tilde{a}^{ - \gamma }     \notag \\
                                                                              & \quad \times \frac{   (1- \Psi_{\alpha})    \tilde{a}^{ - \gamma }    + (\alpha-2)\Psi_{\alpha}             }{   \left [    (1- \Psi_{\alpha})   \tilde{a}^{ - \gamma }  + \Psi_{\alpha}         \right ]^{2}      }      .
\label{eq:d2SBH2SBH0_power}
\end{align}
Equation\ (\ref{eq:d2SBH2SBH0_power}) implies that $ \ddot{S}_{\rm{BH}} < 0$ should be satisfied in the last stage when $\alpha < 2$.
The details of the calculation are summarized in Appendix \ref{SL}.

\begin{figure} [t] 
\begin{minipage}{0.495\textwidth}
\begin{center}
\scalebox{0.33}{\includegraphics{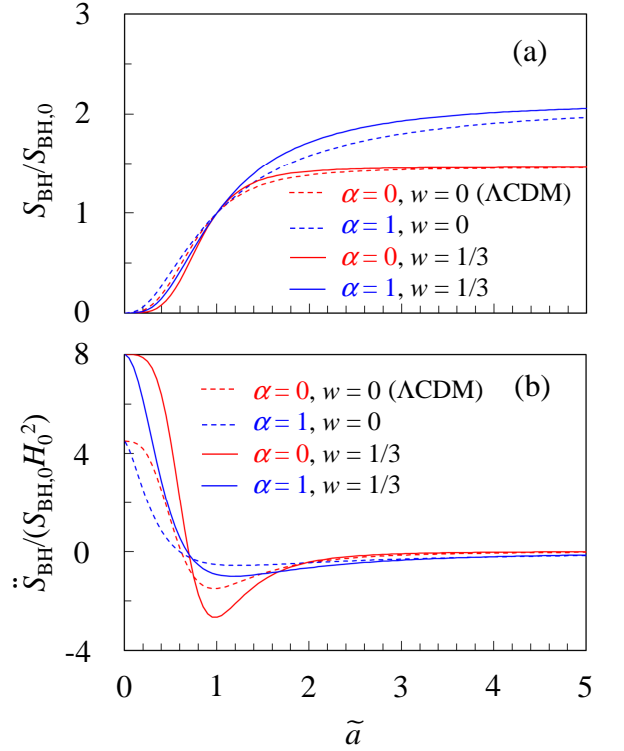}}
\end{center}
\end{minipage}
\caption{Evolution of the normalized entropic parameters for the present model for $\Psi_{\alpha} =0.685$. 
(a) $S_{\rm{BH}} / S_{\rm{BH},0}$.
(b) $\ddot{S}_{\rm{BH}}/ (S_{\rm{BH},0} H_{0}^{2})$.
The dashed and solid lines represent $w=0$ and $w=1/3$, respectively.
The red and blue lines represent $\alpha =0$ and $\alpha =1$, respectively.
}
\label{Fig-S-a}
\end{figure}

\begin{figure} [t] 
\begin{minipage}{0.495\textwidth}
\begin{center}
\scalebox{0.34}{\includegraphics{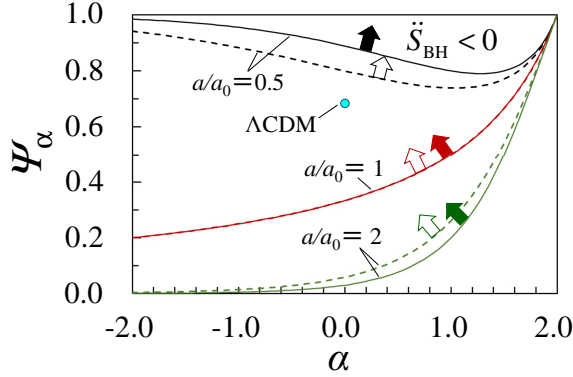}}
\end{center}
\end{minipage}
\caption{Boundary of $\ddot{S}_{\rm{BH}} = 0$ in the $(\alpha, \Psi_{\alpha})$ plane for various values of $a/a_{0}$.
The dashed and solid lines represent $w=0$ and $w=1/3$, respectively.
The arrow attached to each boundary indicates the relaxation-process-side region that satisfies $\ddot{S}_{\rm{BH}} < 0$.
The open circle represents $(\alpha, \Psi_{\alpha}) = (0, 0.685)$ for the $\Lambda$CDM model. 
When $a/a_{0} =1$, the boundary for $w=0$ is the same as that for $w=1/3$, where the boundary is given by $\Psi_{\alpha} = 1/(3 - \alpha)$.
 }
\label{Fig-dS2dt2_plane_power}
\end{figure}

We now observe the evolution of $S_{\rm{BH}}$ and $\ddot{S}_{\rm{BH}}$ for the present model for both $w=0$ and $w=1/3$.
To examine typical results, $\alpha$ is set to $0$ and $1$, and $\Psi_{\alpha}$ is set to $0.685$.
As shown in Fig.\ \ref{Fig-S-a}(a), $S_{\rm{BH}}$ increases with $\tilde{a}$.
That is, the second law of thermodynamics, $\dot{S}_{\rm{BH}} \geq 0$, is satisfied in both cases.
In addition, $S_{\rm{BH}}$ approaches a positive value that depends on $\alpha$ and $\Psi_{\alpha}$ but not on $w$.
The positive value is given by $S_{\rm{BH}} / S_{\rm{BH},0} = \Psi_{\alpha}^{-2/(2-\alpha)}$, which is obtained by applying $ \tilde{a} \rightarrow \infty$ to Eq.\ (\ref{eq:SBH_power}) with $\alpha < 2$.
In fact, $S_{\rm{BH}}$ rapidly increases in the early stage and gradually approaches a positive value in the last stage.
Consequently, $\ddot{S}_{\rm{BH}}$ is positive in the early stage and negative in the last stage, as shown in Fig.\ \ref{Fig-S-a}(b).
These results are consistent with those in Ref.\ \cite{Koma14}.
That is, in both cases, maximization of entropy, $\ddot{S}_{\rm{BH}} < 0$, should be satisfied in the last stage.
Of course, the entropic parameters for $w=1/3$ are quantitatively different from those for $w=0$.
However, the evolution of those for $w=1/3$ is similar to that for $w=0$, as for the case of $H/H_{0}$.

As examined above, the universe observed here approaches a kind of equilibrium state in the last stage.
The evolution of the universe is considered to be a relaxation process.
To study the relaxation process systematically, the boundary required for $\ddot{S}_{\rm{BH}} = 0$ is calculated.
(The boundary of $\ddot{S}_{\rm{BH}} = 0$ for $w=0$ was discussed in Ref.\ \cite{Koma14}.)
Using Eq.\ (\ref{eq:d2Sdt2_power_1B}) and assuming $\dot{H} \neq 0$, we obtain the boundary of $\ddot{S}_{\rm{BH}} = 0$, which is given by
\begin{align}  
   \Psi_{\alpha} = \frac{1}{3 - \alpha }    \left (  \frac{  H }{  H_{0} } \right )^{2-\alpha}   .
\label{eq:d2Sdt2 is 0}
\end{align}
When $\alpha = 2$, $\Psi_{\alpha} = 1$ is obtained from this equation.
When $\alpha \neq 2$, substituting Eq.\ (\ref{eq:Sol_HH0_power}) into Eq.\ (\ref{eq:d2Sdt2 is 0}) yields
\begin{align}  
   \Psi_{\alpha} = \frac{1}{3 - \alpha } \left [ (1- \Psi_{\alpha})  \tilde{a}^{- \gamma} + \Psi_{\alpha} \right ] . 
\label{eq:d2Sdt2_power_0_aa0}
\end{align}
Solving Eq.\ (\ref{eq:d2Sdt2_power_0_aa0}) with respect to $\Psi_{\alpha}$ yields the following boundary required for $\ddot{S}_{\rm{BH}} = 0$:
\begin{align}  
   \Psi_{\alpha} =  \frac{ \tilde{a}^{- \gamma}     }{  2 - \alpha +  \tilde{a}^{- \gamma}     }  ,
\label{eq:d2Sdt2_power_0_aa0_2}
\end{align}
where $\gamma = \frac{3 (1+w) (2-\alpha)}{2}$ from Eq.\ (\ref{aa0gama}).
The influence of $w$ is included in $\gamma$.

Using Eq.\ (\ref{eq:d2Sdt2_power_0_aa0_2}), the boundary of $\ddot{S}_{\rm{BH}} = 0$ for various values of $a/a_{0}$ can be plotted on the $(\alpha, \Psi_{\alpha})$ plane.
In Fig.\ \ref{Fig-dS2dt2_plane_power}, $a/a_{0}$ is set to $0.5$, $1$, and $2$, to observe typical boundaries.
In this figure, $a/a_{0}$ represents the normalized scale factor.
The dashed and solid lines represent $w=0$ and $w=1/3$, respectively.
The arrow attached to each boundary indicates the relaxation-process-side region that satisfies $\ddot{S}_{\rm{BH}} < 0$.
For each boundary, the upper side corresponds to this region.
For both $w=0$ and $w=1/3$, this region gradually extends downward with increasing $a/a_{0}$.
When $a/a_{0} =1$, the boundary for $w=0$ is the same as that for $w=1/3$, where the boundary is given by $\Psi_{\alpha} = 1/(3 - \alpha)$ from Eq.\ (\ref{eq:d2Sdt2_power_0_aa0_2}).
In both cases, a small-$\alpha$ and large-$\Psi_{\alpha}$ region tends to satisfy $\ddot{S}_{\rm{BH}} < 0$ at the present time.
The properties of the boundary of $\ddot{S}_{\rm{BH}} = 0$ for both cases are similar to each other.
Note that the boundary for $w=1/3$ is quantitatively different from that for $w=0$, except when $a/a_{0} =1$.

\subsection{Dynamical temperature $T_{H}$ on the horizon} 
\label{Temperature for the present model}

\begin{figure} [b] 
\begin{minipage}{0.495\textwidth}
\begin{center}
\scalebox{0.33}{\includegraphics{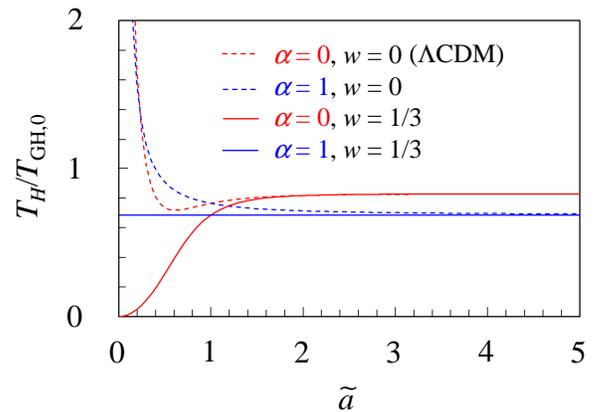}}
\end{center}
\end{minipage}
\caption{Evolution of the normalized $T_{H}$ for the present model for $\Psi_{\alpha} =0.685$. 
The dashed and solid lines represent $w=0$ and $w=1/3$, respectively.
The red and blue lines represent $\alpha =0$ and $\alpha =1$, respectively.
The horizontal straight line corresponds to [$\alpha=1$, $w=1/3$].
}
\label{Fig-T-a}
\end{figure}

\begin{figure*} [htb] 
 \begin{minipage}{0.495\hsize}
  \begin{center}
   \scalebox{0.3}{\includegraphics{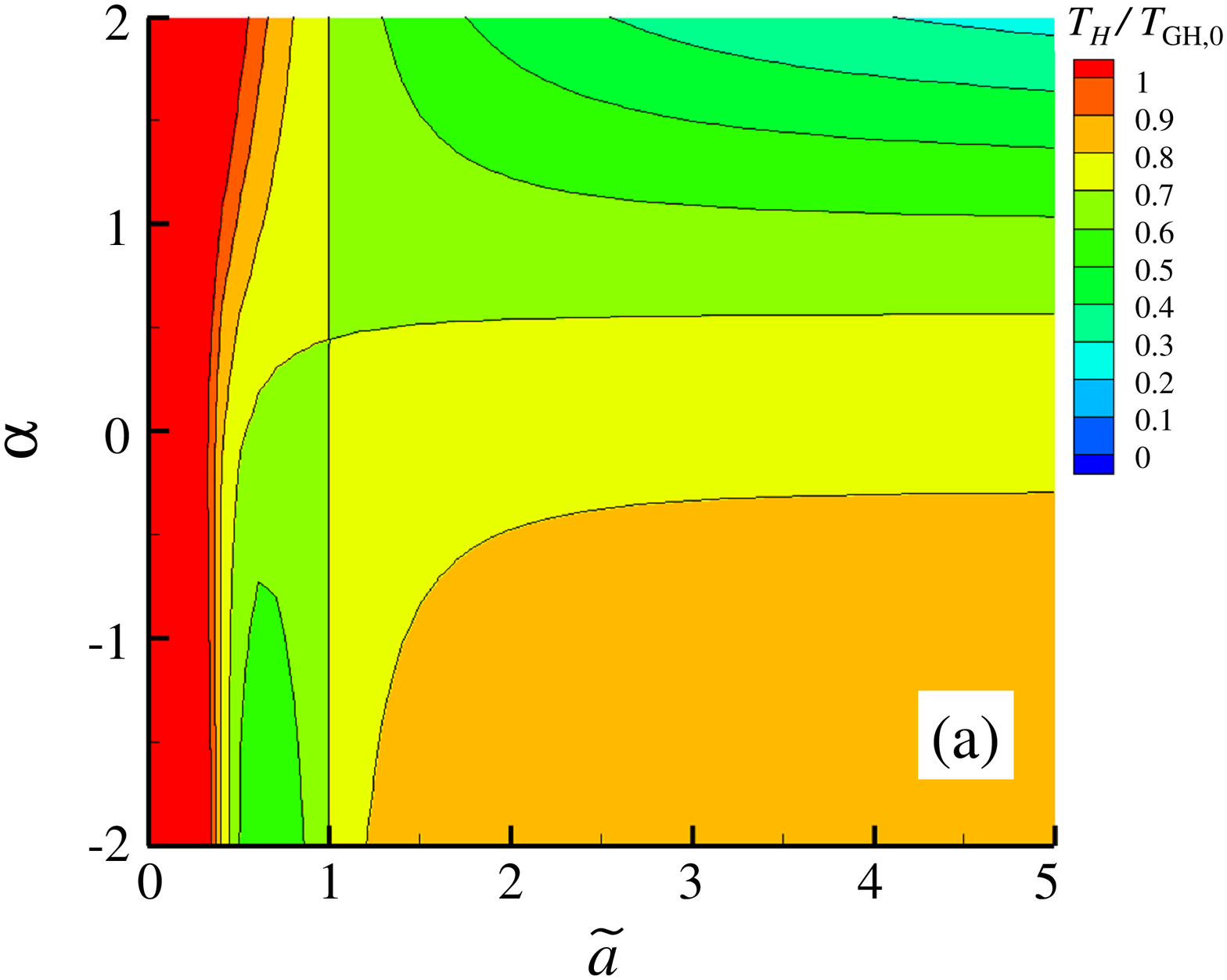}}\\  
  \end{center}
 \end{minipage}
 \begin{minipage}{0.495\hsize}
  \begin{center}
   \scalebox{0.3}{\includegraphics{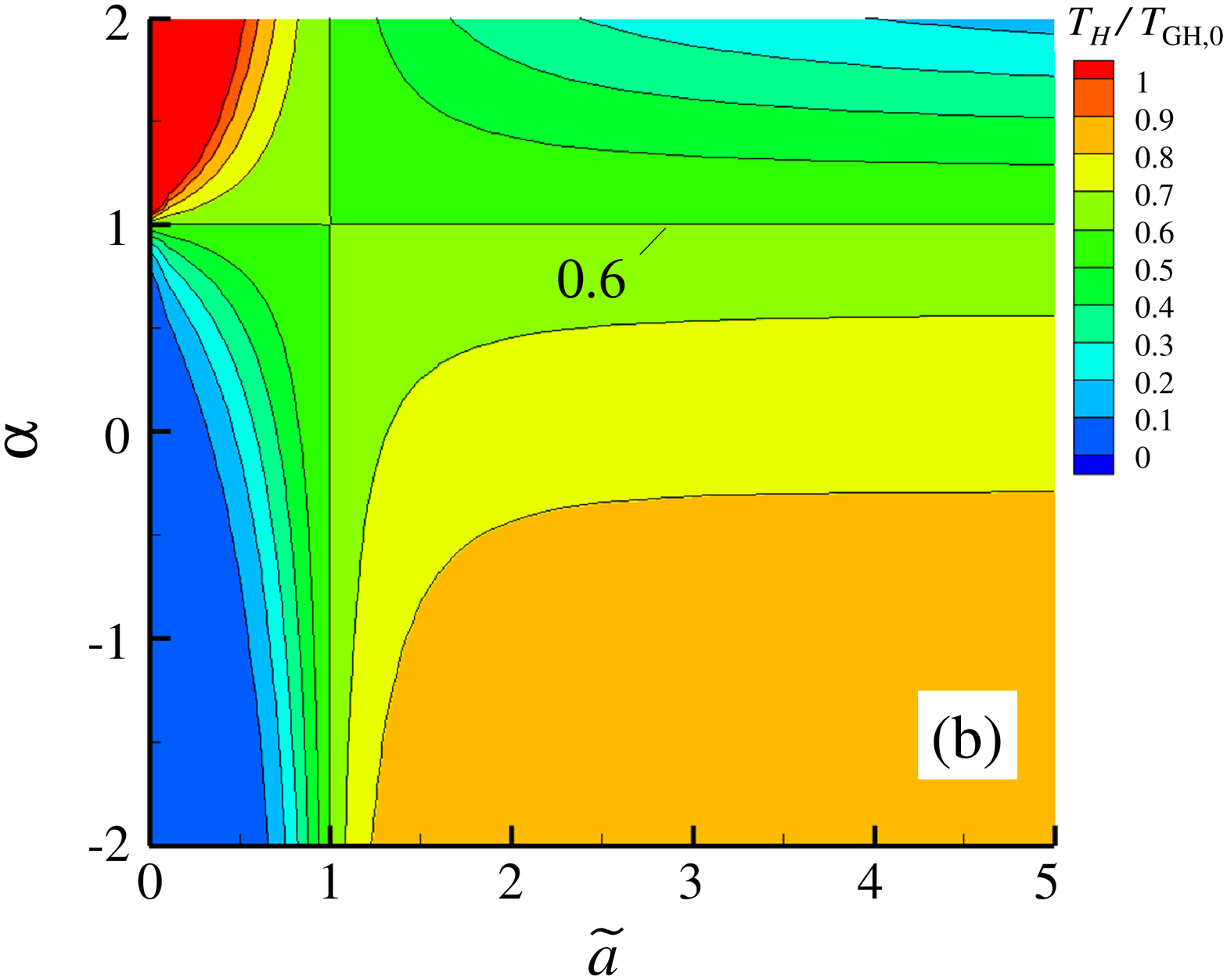}}\\
  \end{center}
 \end{minipage}
\caption{Contours of the normalized $T_{H}$ in the $(\tilde{a}, \alpha)$ plane for the present model for $\Psi_{\alpha} = 0.6$. 
(a) $w=0$. (b) $w=1/3$.
The horizontal axis represents the normalized scale factor $\tilde{a}$, which increases with time.
In (b), the horizontal straight contour line at $\alpha=1$ corresponds to $T_{H}/T_{\rm{GH},0} = \Psi_{\alpha} = 0.6$, which is given by Eq.\ (\ref{TH-TGH0_cst_w13alpha1}).
Note that, in (a) and (b), the vertical straight lines at $\tilde{a}=1$ corresponds to Eq.\ (\ref{TH-TGH0_cst_a=1}).
 }
\label{Fig-a-alpha_T_(w0-w13)_Map}
\end{figure*}

The evolution of the parameters examined, such as $H/H_{0}$, $q$, $S_{\rm{BH}}$, and $\ddot{S}_{\rm{BH}}$), for the present model, was found to be similar for both $w=0$ and $w=1/3$.
However, we expect that the evolution of the dynamical temperature for $w=1/3$ is different from that for $w=0$.
In this subsection, we therefore examine the dynamical temperature $T_{H}$ for the present model.

Substituting Eq.\ (\ref{eq:Back_power_1}) into Eq.\ (\ref{eq:T_H_mod_0}) yields 
\begin{align}
\frac{T_{H}}{ T_{\rm{GH},0} }  &= \frac{H}{H_{0}}  \left ( 1 + \frac{ \dot{H} }{ 2 H^{2} }\right ) \notag \\
                                         &= \frac{H}{H_{0}}  \left ( 1 - \frac{3(1+w)}{4}   \left (    1 -   \Psi_{\alpha} \left (  \frac{H}{H_{0}} \right )^{\alpha -2}  \right ) \right ) , 
\label{TH-TGH0}
\end{align}
where $T_{\rm{GH},0}$ is the Gibbons--Hawking temperature at the present time, given by $\hbar  H_{0} /(2 \pi  k_{B})$.
The normalized $T_{H}$ is not negative in the present model, because $w=0$, $w=1/3$, $0 \leq \Psi_{\alpha} \leq 1$, and $H > 0$ are considered.
Substituting Eq.\ (\ref{eq:Sol_HH0_power}) into Eq.\ (\ref{TH-TGH0}) and performing several calculations yields 
\begin{align}
\frac{T_{H}}{ T_{\rm{GH},0} } &=  \left [ (1- \Psi_{\alpha})   \tilde{a}^{ - \gamma}  + \Psi_{\alpha}  \right ]^{\frac{1}{2-\alpha}} \notag \\
                                & \quad \times  \left [ 1 - \frac{3(1+w)}{4}   \left (    1 -   \frac{  \Psi_{\alpha} }{ (1- \Psi_{\alpha})  \tilde{a}^{ - \gamma}   + \Psi_{\alpha}   } \right ) \right ]  \notag \\
                                &=  \frac{(1-3w) (1- \Psi_{\alpha})  \tilde{a}^{ - \gamma}    + 4 \Psi_{\alpha}   }{4  \left [ (1- \Psi_{\alpha})   \tilde{a}^{ - \gamma}  + \Psi_{\alpha}  \right ]^{\frac{1-\alpha}{2-\alpha}} } ,    \notag \\
\label{TH-TGH0_2}
\end{align}
where $\gamma$ is $\frac{3 (1+w) (2-\alpha)}{2}$ given by Eq.\ (\ref{aa0gama}).
Also, Eq.\ (\ref{TH-TGH0_2}) includes a coefficient $(1-3w)$, which affects the properties of $T_{H}$.
For example, substituting $w=1/3$ and $\alpha =1$ into Eq.\ (\ref{TH-TGH0_2}) yields
\begin{align}
 \frac{T_{H}}{ T_{\rm{GH},0} }=  \Psi_{\alpha}   \quad  (\textrm{for} \quad w=\frac{1}{3} \quad \textrm{and} \quad \alpha =1).
\label{TH-TGH0_cst_w13alpha1}
\end{align}
The obtained temperature does not depend on $\tilde{a}$.

Using Eq.\ (\ref{TH-TGH0_2}), we study two specific cases: $\tilde{a}=1$ and $\tilde{a} \rightarrow \infty$.
Firstly, substituting $\tilde{a}=1$ into Eq.\ (\ref{TH-TGH0_2}) yields
\begin{align}
 \frac{T_{H}}{ T_{\rm{GH},0} } =  \frac{(1-3w) (1- \Psi_{\alpha}) + 4 \Psi_{\alpha}   }{4}     \quad  (\textrm{for} \quad \tilde{a} =1) .
\label{TH-TGH0_cst_a=1}
\end{align}
Equation\ (\ref{TH-TGH0_cst_a=1}) indicates that $T_{H}$ does not depend on $\alpha$ at the present time.
For $w = 0$, Eq.\ (\ref{TH-TGH0_cst_a=1}) is written as $T_{H} / T_{\rm{GH},0} =  (1 + 3 \Psi_{\alpha})/4$.
For $w = 1/3$, Eq.\ (\ref{TH-TGH0_cst_a=1}) is written as $T_{H} / T_{\rm{GH},0} =  \Psi_{\alpha}$, which is equivalent to Eq.\ (\ref{TH-TGH0_cst_w13alpha1}).
Secondly, substituting $\tilde{a} \rightarrow \infty$ into Eq.\ (\ref{TH-TGH0_2}) with $\alpha < 2$ yields
\begin{align}
 \frac{T_{H}}{ T_{\rm{GH},0} } =  \Psi_{\alpha}^{\frac{1}{2-\alpha}}  \quad  (\textrm{for} \quad \tilde{a}  \rightarrow \infty)  . 
\label{TH-TGH0_cst_a=inf}
\end{align}
Equation\ (\ref{TH-TGH0_cst_a=inf}) indicates that $T_{H}$ does not depend on $w$ when $\tilde{a} \rightarrow \infty$.

We now observe the evolution of the normalized $T_{H}$ for the present model for both $w=0$ and $w=1/3$.
To examine typical results, $\alpha$ is set to $0$ and $1$, and $\Psi_{\alpha}$ is set to $0.685$, equivalent to $\Omega_{\Lambda}$ for the $\Lambda$CDM model.  

As shown in Fig.\ \ref{Fig-T-a}, when $\tilde{a} \lessapprox 0.6$, $T_{H}$ for $w=0$ decreases with $\tilde{a}$, whereas $T_{H}$ for $w=1/3$ does not decrease.
The evolution of $T_{H}$ for $w=1/3$ is different from that for $w=0$ in the very early stage.
In the last stage, $T_{H}$ gradually approaches a positive value, $T_{H}/T_{\rm{GH},0} = \Psi_{\alpha}^{1/(2-\alpha)}$, given by Eq.\ (\ref{TH-TGH0_cst_a=inf}), that depends on $\alpha$ and $\Psi_{\alpha}$ but not on $w$.
In particular, $T_{H}$ for [$\alpha=1$, $w=1/3$] is constant during the evolution of the universe.
The universe at constant $T_{H}$ is not a de Sitter universe because $H/H_{0}$ for [$\alpha=1$, $w=1/3$] varies with $\tilde{a}$, as shown in Fig.\ \ref{Fig-H-a}(a).
We note that $H/H_{0}$ is equivalent to the normalized Gibbons--Hawking temperature because $T_{\rm{GH}} = \hbar H/(2 \pi  k_{B})$.

Figure\ \ref{Fig-T-a} indicates that a universe at constant $T_{H}$ on a dynamic horizon is obtained from the present model for [$\alpha=1$, $w=1/3$].
To observe this from a different viewpoint, we plot contours of the normalized $T_{H}$ in the $(\tilde{a}, \alpha)$ plane.
As shown in Fig.\ \ref{Fig-a-alpha_T_(w0-w13)_Map}, the contour lines are plotted at increments of $0.1$.
We set $\Psi_{\alpha} = 0.6$, to make a certain contour line for $w=1/3$ clear,
as discussed below.

As shown in Fig.\ \ref{Fig-a-alpha_T_(w0-w13)_Map} (a), for $w=0$, the normalized $T_{H}$ for all $\alpha$ varies with $\tilde{a}$.
In contrast, for $w=1/3$, the normalized $T_{H}$ for $\alpha=1$ is indicated by the horizontal straight-contour-line [Fig.\ \ref{Fig-a-alpha_T_(w0-w13)_Map} (b)].
The horizontal straight-contour-line corresponds to $T_{H}/T_{\rm{GH},0} = \Psi_{\alpha} =0.6$, which is given by Eq.\ (\ref{TH-TGH0_cst_w13alpha1}).
Also, in the early stage ($\tilde{a} \ll 1$), the normalized $T_{H}$ for $\alpha >1 $ is high, whereas the normalized $T_{H}$ for $ \alpha <1 $ is low.
These results indicate that $\alpha=1$ can be considered a kind of critical value when the normalized $T_{H}$ for $w=1/3$ is discussed in the present model.

In this way, the evolution of the normalized $T_{H}$ for $w=1/3$ is different from that for $w=0$.
In addition, we can obtain a universe at constant $T_{H}$ on a dynamic horizon from the present model for [$\alpha=1$, $w=1/3$].
The obtained universe corresponds to a radiation-dominated universe that includes an extra driving term proportional to $H$.
The universe is expected to be a good model for studying relaxation processes for the universe at constant $T_{H}$ because systems at constant temperature play important roles in thermodynamics and statistical physics.
In the next section, we examine specific conditions required for constant $T_{H}$, based on the definition of the dynamical temperature.

\section{Constant $T_{H}$ model} 
\label{Constant T}

In this section, we examine the conditions required for a constant $T_{H}$ and formulate a cosmological model that can describe a universe at constant $T_{H}$ on a dynamic horizon, and then discuss the properties of the constant $T_{H}$ universe of the formulated model.
The universe considered in this section should be different from the late Universe,
but should help in studying the relaxation processes for thermodynamic quantities on a dynamic horizon.

From Eq.\ (\ref{eq:T_H_mod}), the temperature on the horizon of a flat FRW universe is written as
\begin{equation}
 T_{H} = \frac{ \hbar H}{   2 \pi  k_{B}  }  \left ( 1 + \frac{ \dot{H} }{ 2 H^{2} }\right ) .
\label{eq:T_H_mod2}
\end{equation}
To examine the conditions required for constant $T_{H}$, we consider a non-dimensional parameter $\psi$, written as
\begin{equation}
\psi  =  \frac{H}{H_{0}} \left ( 1 + \frac{ \dot{H} }{ 2 H^{2} }\right ) , 
\label{psi}
\end{equation}
where $\psi$ is assumed to be constant.
When this equation is satisfied, $T_{H}$ is constant because $T_{H}$ can be written as
\begin{align}
   T_{H} &= \frac{ \hbar  \psi H_{0}}{   2 \pi  k_{B}  }  \quad (\textrm{for constant $T_{H}$})  ,
\label{cstTH}
\end{align}
where Eqs.\ (\ref{eq:T_H_mod2}) and (\ref{psi}) are used.
The above equation indicates that $T_{H}$ is proportional to $\psi$.
Note that $\psi$ should also be related to surface gravity because $T_{H}$ given by Eq.\ (\ref{eq:T_H_mod2}) is based on the relationship between the temperature and the surface gravity \cite{Dynamical-T-20072014,Tu2018,Tu2019,ApparentHorizon2022,Mathew2023}.

Solving Eq.\ (\ref{psi}) with respect to $\dot{H}$ yields  
\begin{align}
    \dot{H} &= - 2 H^{2}  +  2 \psi H_{0}  H  .
\label{Cosmo_T_H_mod_cst1}
\end{align}
We expect that Eq.\ (\ref{Cosmo_T_H_mod_cst1}) is related to cosmological models.
Based on this expectation, we attempt to formulate a cosmological model that satisfies Eq.\ (\ref{Cosmo_T_H_mod_cst1}).
To this end, we consider the present model again.
From Eq.\ (\ref{eq:Back_power_1}), the cosmological equation for the present model is written as
\begin{align}
    \dot{H} &= - \frac{3}{2} (1+w)  H^{2}  +  \frac{3}{2}   (1+w)  \Psi_{\alpha} H_{0}^{2} \left (  \frac{H}{H_{0}} \right )^{\alpha}  .
\label{eq:Back_power_2}
\end{align}
By comparing Eqs.\ (\ref{Cosmo_T_H_mod_cst1}) and (\ref{eq:Back_power_2}), we find 
\begin{equation}
   \alpha  =1     \quad   \textrm{and}  \quad  w= \frac{1}{3}  .
\label{w_alpha_T_H_mod_cst}
\end{equation}
In fact, substituting Eq.\ (\ref{w_alpha_T_H_mod_cst}) into Eq.\ (\ref{eq:Back_power_2}) yields
\begin{align}
    \dot{H} &=  - 2 H^{2}  +  2 \Psi_{\alpha} H_{0}  H  .
\label{eq:Back_power_2_w13a1}
\end{align}
This equation is equivalent to Eq.\ (\ref{Cosmo_T_H_mod_cst1}) for $\Psi_{\alpha} = \psi$.
The above cosmological model, hereafter `the constant $T_{H}$ model', can describe a universe at constant $T_{H}$ on a dynamic horizon.
The constant $T_{H}$ model corresponds to the present model for [$\alpha  =1$,  $w= 1/3$].
The model obtained here is a viable scenario
in that other models can also satisfy Eq.\ (\ref{Cosmo_T_H_mod_cst1}).
For example, substituting $w=0$ and $3 f_{\Lambda}(t)/2 = - H^{2}/2 +2 \psi H_{0} H$ into Eq.\ (\ref{eq:Back_f}) can yield Eq.\ (\ref{Cosmo_T_H_mod_cst1}).
Even in this case, the background evolution of the universe is equivalent to that for the constant $T_{H}$ model because Eq.\ (\ref{Cosmo_T_H_mod_cst1}) is satisfied.
In the present study, we use the constant $T_{H}$ model as a viable scenario.

\begin{figure} [t] 
\begin{minipage}{0.495\textwidth}
\begin{center}
\scalebox{0.33}{\includegraphics{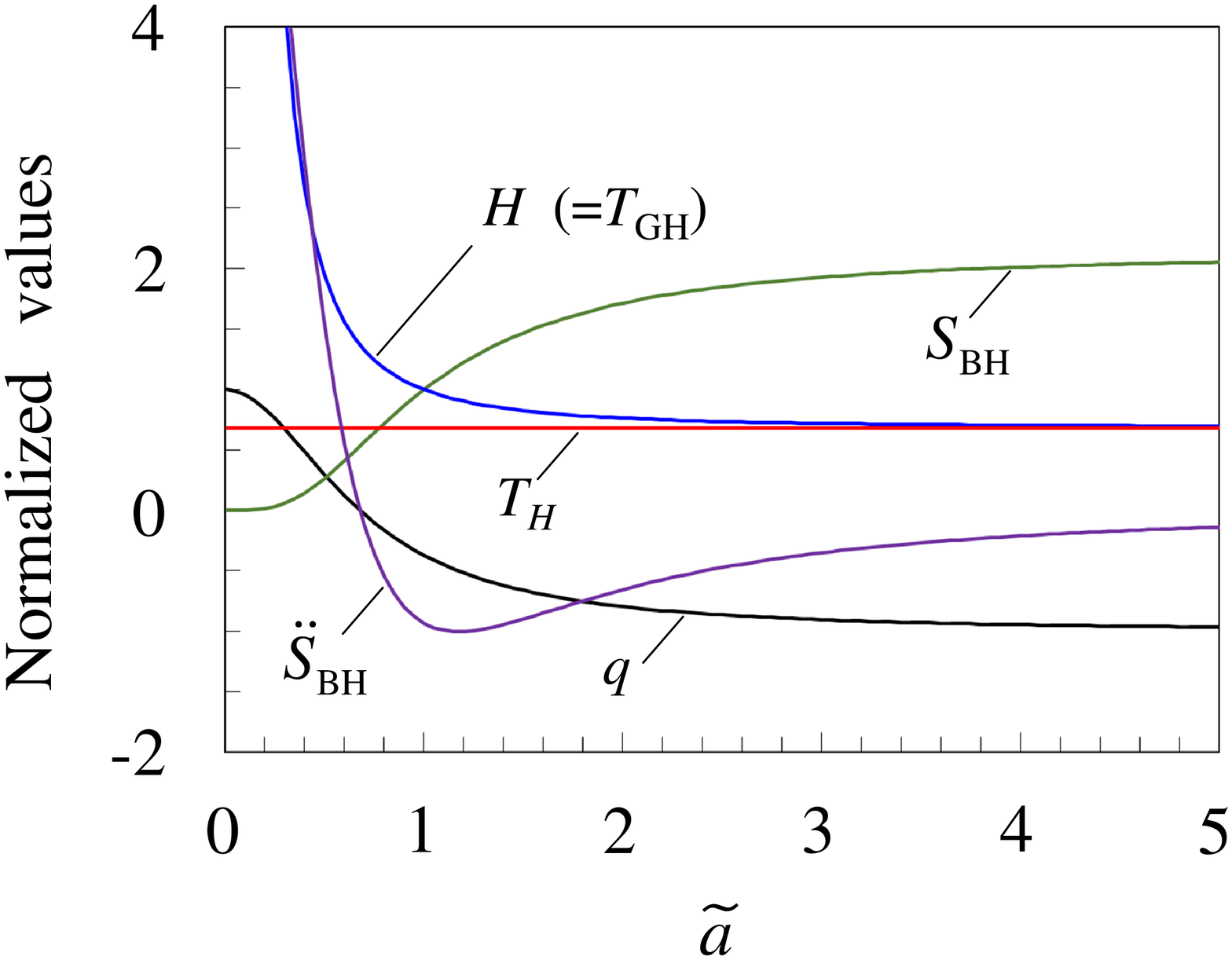}}
\end{center}
\end{minipage}
\caption{Evolution of normalized parameters for the constant $T_{H}$ model for $\Psi_{\alpha} =0.685$. 
The parameters are replotted and $q$ is not normalized.
The normalized $H$ is equivalent to the normalized $T_{\rm{GH}}$ (see the text.).
The constant $T_{H}$ model corresponds to the present model for [$\alpha  =1$, $w= 1/3$].
}
\label{Fig-HqS-a_cstT}
\end{figure}

As mentioned above, $\psi$ is considered to be related to the horizon temperature and the surface gravity.
Also, $\Psi_{\alpha}$ is a kind of density parameter for the effective dark energy.
Therefore, $\Psi_{\alpha} = \psi$ may imply that the effective dark energy is related to the temperature and the surface gravity.
In this study, we accept this relation and assume $\Psi_{\alpha} = \psi$.
Consequently, from Eq.\ (\ref{cstTH}), the constant normalized temperature is written as
\begin{align}
  \frac{T_{H}} {T_{\rm{GH},0} } = \frac{  \frac{ \hbar  \psi H_{0}}{   2 \pi  k_{B}  } }{ \frac{ \hbar  H_{0}}{   2 \pi  k_{B}  } }=  \psi  = \Psi_{\alpha}   ,
\label{cstTH_TGH_Psi}
\end{align}
where $T_{\rm{GH},0}$ is the Gibbons--Hawking temperature at the present time, given by $\hbar  H_{0} /(2 \pi  k_{B})$.

We now observe the evolution of several parameters for the constant $T_{H}$ model and examine the relaxation processes for the universe.
To observe typical results, $\Psi_{\alpha}$ is set to $0.685$, as in previous sections.

\begin{figure} [t] 
\begin{minipage}{0.495\textwidth}
\begin{center}
\scalebox{0.33}{\includegraphics{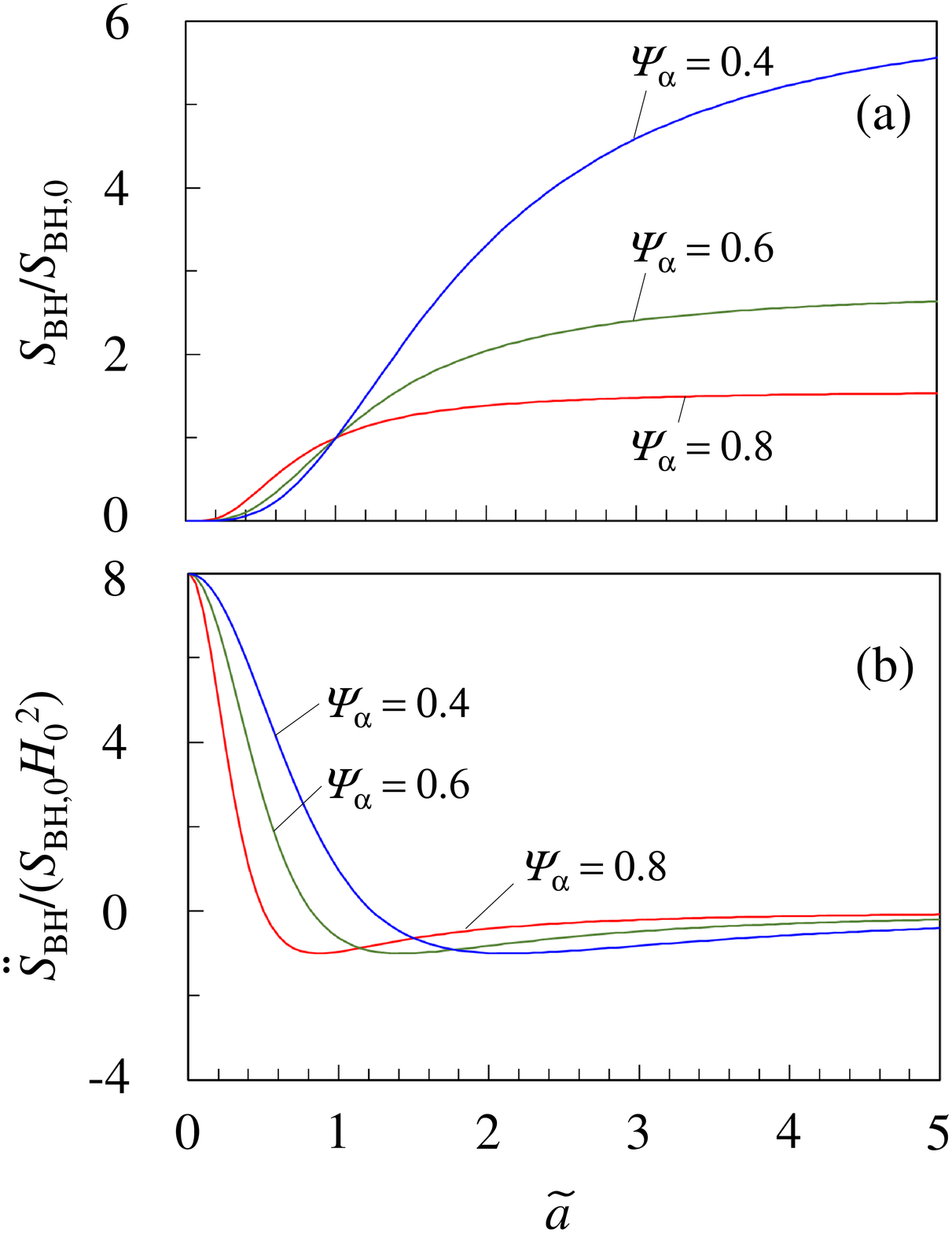}}
\end{center}
\end{minipage}
\caption{Evolution of normalized entropic parameters for the constant $T_{H}$ model for various values of $\Psi_{\alpha}$. 
(a) $S_{\rm{BH}} / S_{\rm{BH},0}$.
(b) $\ddot{S}_{\rm{BH}}/ (S_{\rm{BH},0} H_{0}^{2})$.
The constant $T_{H}$ model corresponds to the present model for [$\alpha  =1$, $w= 1/3$].
}
\label{Fig-S-a_Tcst_psi}
\end{figure}

\begin{figure} [t] 
\begin{minipage}{0.495\textwidth}
\begin{center}
\scalebox{0.30}{\includegraphics{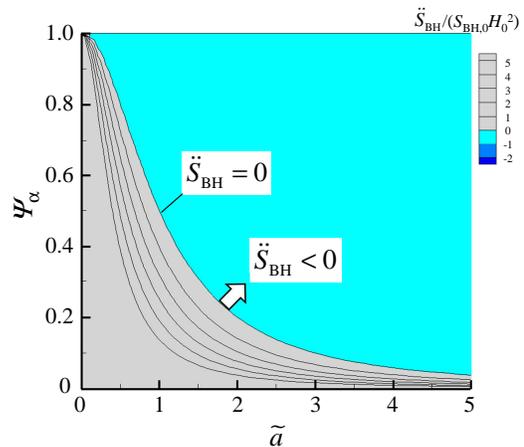}}
\end{center}
\end{minipage}
\caption{Contours of normalized $\ddot{S}_{\rm{BH}}$ in the $(\tilde{a}, \Psi_{\alpha})$ plane for the constant $T_{H}$ model. 
The vertical axis $\Psi_{\alpha}$ is equivalent to the normalized $T_{H}$.
The arrow attached to the line of $\ddot{S}_{\rm{BH}} = 0$ indicates a region that satisfies $\ddot{S}_{\rm{BH}} < 0$.
Unsatisfied regions are displayed in gray, to make the boundary of $\ddot{S}_{\rm{BH}} = 0$ clear.
The contour lines are plotted at increments of $1$.
The color scale bar is based on the normalized value, which is calculated from Eq.\ (\ref{eq:d2SBH2SBH0_power}), applying $\alpha  =1$ and $w= 1/3$.
 }
\label{Fig-a-Psi_d2tdt2_Map}
\end{figure}

As shown in Fig.\ \ref{Fig-HqS-a_cstT}, $T_{H}$ is constant during the evolution of the universe.
The value of the normalized $T_{H}$ is $0.685$ from Eq.\ (\ref{cstTH_TGH_Psi}).
The other parameters (namely $H$, $q$, $S_{\rm{BH}}$, and $\ddot{S}_{\rm{BH}}$) gradually approach a constant value in the last stage.
The final state corresponds to a de Sitter universe whose temperature is $T_{H}$.

We note that the Gibbons--Hawking temperature $T_{\rm{GH}} = \hbar H/(2 \pi  k_{B})$ is proportional to $H$, as given by Eq.\ (\ref{eq:T_H1}).
Therefore, the normalized $T_{\rm{GH}}$, namely $T_{\rm{GH}}/T_{\rm{GH},0}$, is $H/H_{0}$ and is equivalent to the normalized $H$.
Thus, Fig.\ \ref{Fig-HqS-a_cstT} indicates that the normalized $T_{\rm{GH}}$ decreases with $\tilde{a}$ and gradually approaches the normalized constant $T_{H}$.

The evolution of these parameters can be interpreted as a relaxation process at constant $T_{H}$.
To discuss the relaxation process, we examine the evolution of $S_{\rm{BH}}$ and $\ddot{S}_{\rm{BH}}$ for various values of $\Psi_{\alpha}$. 
To study typical results, $\Psi_{\alpha}$ is set to $0.4$, $0.6$, and $0.8$.
(Note that $\Psi_{\alpha}$ is equivalent to the value of the normalized $T_{H}$.)
As shown in Fig.\ \ref{Fig-S-a_Tcst_psi}(a), the normalized ${S}_{\rm{BH}}$ increases with $\tilde{a}$ and gradually approaches a positive value that depends on $\Psi_{\alpha}$.
The normalized value is given by $S_{\rm{BH}} / S_{\rm{BH},0} = \Psi_{\alpha}^{-2/(2-\alpha)} = \Psi_{\alpha}^{-2}$, as considered in Sec.\ \ref{Entropy for the present model}, where $\alpha =1$ is used for the constant $T_{H}$ model.
The evolution of $S_{\rm{BH}}$ depends on $\Psi_{\alpha}$; that is, the larger $\Psi_{\alpha}$ is, the earlier ${S}_{\rm{BH}}$ approaches a positive value.
Also, as shown in Fig.\ \ref{Fig-S-a_Tcst_psi}(b), the normalized $\ddot{S}_{\rm{BH}}$ is positive initially and negative finally.
Accordingly, the maximization of entropy, namely $\ddot{S}_{\rm{BH}} < 0$, is satisfied in the last stage.

In the constant $T_{H}$ model, the normalized $T_{H}$ is equivalent to $\Psi_{\alpha}$, as shown in Eq.\ (\ref{cstTH_TGH_Psi})
and hence their influence on the model can be seen as being the same.
Accordingly, to observe the influence of $T_{H}$, we examine a temporal $\Psi_{\alpha}$-region that satisfies the maximization of entropy, using contours of $\ddot{S}_{\rm{BH}}$ in the $(\tilde{a}, \Psi_{\alpha})$ plane. 
In Fig.\ \ref{Fig-a-Psi_d2tdt2_Map}, the arrow attached to the line $\ddot{S}_{\rm{BH}} = 0$ indicates a region that satisfies the maximization of entropy, $\ddot{S}_{\rm{BH}} < 0$.
The line $\ddot{S}_{\rm{BH}} = 0$ is equivalent to the boundary calculated from Eq.\ (\ref{eq:d2Sdt2_power_0_aa0_2}).
As shown in Fig.\ \ref{Fig-a-Psi_d2tdt2_Map}, the normalized $\ddot{S}_{\rm{BH}}$ tends to be positive in the early stage and negative in the last stage.
In addition, the larger $\Psi_{\alpha}$ is, the earlier $\ddot{S}_{\rm{BH}}<0$ is satisfied.
These results imply that the higher $T_{H}$ is, the earlier the entropy should be maximized.

In this way, using the constant $T_{H}$ model, we can examine the relaxation processes for a universe at constant temperature on a dynamic horizon.
Of course, this model is simply one viable scenario with a constant horizon temperature.
The obtained universe is different from the late Universe described by $\Lambda$CDM models because $\alpha  =1$ and $w= 1/3$ are considered here.
However, we expect that the constant $T_{H}$ model will contribute to the study of thermodynamics and statistical physics on dynamic horizons because the horizon temperature is constant in de Sitter universes. 
For example, the holographic equipartition law of energy \cite{Padma2010,ShuGong2011} should be properly applied to the dynamic horizon in a constant $T_{H}$ model.
Based on this, the energy $E_{H}$ is written as $ E_{H} =  N_{H}  \times \frac{1}{2} k_{B} T_{H}  = 2 S_{\rm{BH}}  T_{H}$, where $N_{H}$ is the number of degrees of freedom on the horizon, given by $N_{H} = 4 S_{\rm{BH}} / k_{B}$.
Using these thermodynamic quantities, thermodynamic relations can be examined on the dynamic horizon at constant temperature.
Also, we may discuss the relationship between holographic entanglement entropy \cite{RyuTakayanagi2006,Takayanagi2022,deSitter2,deSitter3} and thermodynamic entropy on the dynamic horizon by extending this model.
Those tasks are left for future research.

\section{Conclusions}
\label{Conclusions}

To clarify the thermodynamics on a dynamic horizon, we examined the evolution of the dynamical temperature $T_{H}$ and the Bekenstein--Hawking entropy $S_{\rm{BH}}$ on the horizon of a flat FRW universe in a $\Lambda(t)$ model.
In this study, we considered a $\Lambda(t)$ model that includes both a power-law term proportional to $H^{\alpha}$ and the equation of state parameter $w$.
Using the present model, we examined a matter-dominated universe ($w=0$) and a radiation-dominated universe ($w=1/3$), setting $\alpha <2$.
Both universes are found to approach de Sitter universes and satisfy maximization of the entropy in the last stage.
The evolution of several parameters (such as $H/H_{0}$, $q$, and $S_{\rm{BH}}$) is similar for $w=0$ and $w=1/3$.
However, the evolution of $T_{H}$ is different for $w=0$ and $w=1/3$.
In particular, $T_{H}$ is constant for $w=1/3$ with $\alpha=1$, although the Hubble parameter $H$ and the Hubble radius $r_{H}$ vary with time, unlike for a de Sitter universe.

To discuss this particular case, we examined the specific conditions required for constant $T_{H}$.
By applying the condition [$\alpha  =1$, $w= 1/3$] to the present model, we formulated a cosmological model that can describe a universe with constant $T_{H}$ on a dynamic horizon.
The formulated constant $T_{H}$ model implies that the density parameter for the effective dark energy is related to $T_{H}$.
It is found that the higher $T_{H}$ is, the earlier the entropy should be maximized.
Using the constant $T_{H}$ model we can examine the relaxation processes for a universe at constant horizon temperature, as if the dynamic horizon is in contact with a heat bath.

The present results may provide new insights for the discussion of thermodynamics and statistical physics on the cosmological horizon.
Detailed studies are needed and are left for future research.

\appendix

\section{$\dot{S}_{\rm{BH}}$ and $\ddot{S}_{\rm{BH}}$ for the $\Lambda(t)$ model with a power-law term} 
\label{SL}

In this section, we calculate $\dot{S}_{\rm{BH}}$ and $\ddot{S}_{\rm{BH}}$ for a $\Lambda(t)$ model that includes both a power-law term and the equation of state parameter.
For this, the present model is given again.
From Eq.\ (\ref{eq:Back_power_1}), the differential equation is  
\begin{align}  
    \dot{H} &= - \frac{3}{2} (1+w)  H^{2}  +  \frac{3}{2}   (1+w)  \Psi_{\alpha} H_{0}^{2} \left (  \frac{H}{H_{0}} \right )^{\alpha}         \notag \\
               & = - \frac{3}{2} (1+w) H^{2}  \left (  1 -   \Psi_{\alpha} \left (  \frac{H}{H_{0}} \right )^{\alpha -2} \right )      .  
\label{eq:Back_power_12}
\end{align}
The solution for $\alpha \neq 2$ given by Eq.\ (\ref{eq:Sol_HH0_power}) is written as 
\begin{equation}  
    \left ( \frac{H}{H_{0}} \right )^{2-\alpha}  =   (1- \Psi_{\alpha})   \tilde{a}^{- \gamma} + \Psi_{\alpha}      ,
\label{eq:Sol_HH0_power_22}
\end{equation}
where $\gamma = \frac{3 (1+w) (2-\alpha)}{2}$ from Eq.\ (\ref{aa0gama}).

The following calculations are based on Refs.\ \cite{Koma14,Koma15}.
The results examined in the previous works are slightly extended because the present model includes the equation of state parameter $w$.

To obtain $\dot{S}_{\rm{BH}}$ for the present model, we first calculate the first derivative of $S_{\rm{BH}}$ from Eq.\ (\ref{eq:SBH2}).
Differentiating Eq.\ (\ref{eq:SBH2}) with respect to $t$ yields \cite{Koma11,Koma12}
\begin{align}  
\dot{S}_{\rm{BH}} &=  \frac{d }{dt} S_{\rm{BH}}  = \frac{d}{dt}   \left ( \frac{K}{H^{2}} \right )  =  \frac{-2K \dot{H} }{H^{3}}    .
\label{eq:dSBH_2_3}      
\end{align}
Substituting Eq.\ (\ref{eq:Back_power_12}) into Eq.\ (\ref{eq:dSBH_2_3}) and applying Eq.\ (\ref{eq:Sol_HH0_power_22}) yields
\begin{align}  
\dot{S}_{\rm{BH}}   
&=  \frac{-2K \dot{H} }{H^{3}}  =  \frac{2K}{H_{0}}  \left ( \frac{- \dot{H} }{H^{2}} \right )    \frac{H_{0}}{H}                                    \notag \\
&=  \frac{2K}{H_{0}}  \frac{3}{2} (1+w) \left (  1 -   \Psi_{\alpha} \left (  \frac{H}{H_{0}} \right )^{\alpha -2}  \right )      \frac{H_{0}}{H}     \notag \\
&=  \frac{3K}{H_{0}} (1+w)  \left (  1 -   \frac{ \Psi_{\alpha} }{ (1- \Psi_{\alpha})   \tilde{a}^{- \gamma} + \Psi_{\alpha}    }  \right )       \notag \\
& \quad \times  \left [ (1- \Psi_{\alpha})  \tilde{a}^{- \gamma} + \Psi_{\alpha}  \right ]^{ \frac{1}{\alpha-2} }                                    \notag \\
&=   \frac{3K}{H_{0}} \frac{  (1+w) (1- \Psi_{\alpha})  \tilde{a}^{- \gamma}    }{      \left [ (1- \Psi_{\alpha})  \tilde{a}^{- \gamma} + \Psi_{\alpha}  \right ]^{ \frac{3-\alpha}{2- \alpha} }  } .
\label{eq:dSBH_2_3_power_Complicate}      
\end{align}   
The obtained $\dot{S}_{\rm{BH}}$ includes $\gamma$ and a coefficient $(1+w)$.
Also, $\gamma$ includes the coefficient $(1+w)$.
Except for these points, Eq.\ (\ref{eq:dSBH_2_3_power_Complicate}) is equivalent to that examined in Refs.\ \cite{Koma14,Koma15}.

In this paper, $1+w \ge 0$, $1 - \Psi_{\alpha} \ge 0$, and $\Psi_{\alpha} \ge 0$ are satisfied because $w= 0$, $w=1/3$, and $0 \le \Psi_{\alpha} \le 1$ are considered.
Accordingly, the second law of thermodynamics on the horizon, namely $\dot{S}_{\rm{BH}} \geq 0$, is satisfied in the present model.
The second law of thermodynamics has been examined for $w=0$ in Ref.\ \cite{Koma14}.

Next, we calculate $\ddot{S}_{\rm{BH}}$.
Differentiating Eq.\ (\ref{eq:dSBH_2_3}) with respect to $t$ yields 
\begin{align}
\ddot{S}_{\rm{BH}}   &= \frac{d}{dt} \dot{S}_{\rm{BH}} = \frac{d}{dt} \left ( \frac{-2K \dot{H} }{H^{3}} \right )    
                               =  - 2 K  \left ( \frac{\ddot{H} }{H^{3}} - \frac{3 \dot{H}^{2} }{H^{4}}  \right )                  \notag  \\
                             &=    2 \frac{K}{H^{2}}  \left ( \frac{   3 \dot{H}^{2}  - \ddot{H} H   }{H^{2}} \right )     
                               =    2 S_{\rm{BH}}       \left ( \frac{   3 \dot{H}^{2}   - \ddot{H} H  }{H^{2}}  \right )     .
\label{eq:d2SB_1}      
\end{align}

We now calculate $\ddot{S}_{\rm{BH}}$ for the present model.
For this, we calculate $3 \dot{H}^{2}   - \ddot{H} H$ in Eq.\ (\ref{eq:d2SB_1}) using Eq.\ (\ref{eq:Back_power_12}).
The detailed calculation is summarized in Ref.\ \cite{Koma14}.
Based on the result, $3 \dot{H}^{2}   - \ddot{H} H $ is written as
\begin{align}  
   3 \dot{H}^{2}   - \ddot{H} H     &=  \frac{3}{2} (1+w) ( - \dot{H} ) H^{2}                                                                                             \notag  \\
                                               & \quad \times     \left [ 1 -  \Psi_{\alpha}  (3 - \alpha )    \left (  \frac{  H }{  H_{0} } \right )^{\alpha-2}   \right ]   . 
\label{eq:B1set5_20}
\end{align}
The above equation includes a coefficient $(1+w)$.
Except for this point, Eq.\ (\ref{eq:B1set5_20}) is equivalent to Eq.\ (C7) of Ref.\ \cite{Koma14}.
Substituting Eq.\ (\ref{eq:B1set5_20}) into Eq.\ (\ref{eq:d2SB_1}) yields
\begin{align}  
\ddot{S}_{\rm{BH}}   
                              &=  \frac{ 2 S_{\rm{BH}}  \frac{3}{2}  (1+w)   ( - \dot{H} ) H^{2}   \left [ 1 -  \Psi_{\alpha}  (3 - \alpha )    \left (  \frac{  H }{  H_{0} } \right )^{\alpha-2}   \right ]     }{H^{2}}     \notag \\
                              &=3 S_{\rm{BH}}   (1+w) ( - \dot{H} )   \left [ 1 -  \Psi_{\alpha}  (3 - \alpha )    \left (  \frac{  H }{  H_{0} } \right )^{\alpha-2}   \right ]   ,
\label{eq:d2Sdt2_power_1}
\end{align}
and applying $S_{\rm{BH}} = K/H^{2}$ given by Eq.\ (\ref{eq:SBH2}) yields
\begin{align}  
\ddot{S}_{\rm{BH}}   &=3 K (1+w) \left (  \frac{ - \dot{H} }{H^{2}}  \right )          \left [ 1 -  \Psi_{\alpha}  (3 - \alpha )    \left (  \frac{  H }{  H_{0} } \right )^{\alpha-2}   \right ]   .
\label{eq:d2Sdt2_power_1B}
\end{align}
In addition, substituting Eq.\ (\ref{eq:Back_power_12}) into Eq.\ (\ref{eq:d2Sdt2_power_1B}) and applying Eq.\ (\ref{eq:Sol_HH0_power_22}) to the resultant equation yields
\begin{align}  
 \ddot{S}_{\rm{BH}}   
 &=3 K (1+w) \left (  \frac{ - \dot{H} }{H^{2}}  \right )          \left [ 1 -  \Psi_{\alpha}  (3 - \alpha )    \left (  \frac{  H }{  H_{0} } \right )^{\alpha-2}   \right ]                \notag \\
 &= 3 K (1+w) \times \frac{3}{2} (1+w) \left (  1 -   \Psi_{\alpha} \left (  \frac{H}{H_{0}} \right )^{\alpha -2} \right )                             \notag \\
 & \quad \times            \left [ 1 -  \Psi_{\alpha}  (3 - \alpha )    \left (  \frac{  H }{  H_{0} } \right )^{\alpha-2}   \right ]     \notag \\
 &= \frac{9K}{2} (1+w)^{2} \left (  1 -  \frac{  \Psi_{\alpha}                      }{   (1- \Psi_{\alpha})   \tilde{a}^{- \gamma} + \Psi_{\alpha}         }  \right )    \notag \\
 & \quad \times           \left [  1 -   \frac{  \Psi_{\alpha}  (3 - \alpha )  }{   (1- \Psi_{\alpha})   \tilde{a}^{- \gamma} + \Psi_{\alpha}         }  \right ]   \notag \\
 &= \frac{9K}{2} (1+w)^{2}  (1- \Psi_{\alpha})   \tilde{a}^{- \gamma}  \notag \\
 & \quad \times    \frac{  (1- \Psi_{\alpha})   \tilde{a}^{- \gamma}   +   (\alpha -2)  \Psi_{\alpha}        }{  \left [ (1- \Psi_{\alpha})  \tilde{a}^{- \gamma}  + \Psi_{\alpha}   \right ]^{2}         }    ,
\label{eq:d2Sdt2_power_1_Complicate}
\end{align}
where $\gamma = \frac{3 (1+w) (2-\alpha)}{2}$ from Eq.\ (\ref{aa0gama}).
Equation\ (\ref{eq:d2Sdt2_power_1_Complicate}) includes a coefficient $(1+w)^2$ and $\gamma$.
Also, $\gamma$ includes $(1+w)$.
Except for these points, Eq.\ (\ref{eq:d2Sdt2_power_1_Complicate}) is equivalent to that examined in Refs.\ \cite{Koma14,Koma15}.
The maximization of the entropy for $w=0$ was discussed in previous works and it was reported that $ \ddot{S}_{\rm{BH}} < 0$ should be satisfied in the last stage when $\alpha < 2$  \cite{Koma14,Koma15}.
In the previous works, $w= 0$ is considered and, therefore, $(1+w)$ is positive.
In this study, similarly, $(1+w)$ is positive because $w= 0$ and $w=1/3$ are considered.
Accordingly, the result reported in Refs.\ \cite{Koma14,Koma15} can also be applied to the present model.
That is, Eq.\ (\ref{eq:d2Sdt2_power_1_Complicate}) indicates that $ \ddot{S}_{\rm{BH}} < 0$ should be satisfied in the last stage when $\alpha < 2$.
We note that $\dot{S}_{\rm{BH}}$ and $\ddot{S}_{\rm{BH}}$ for $\alpha \neq 2$ reduce to those for $\alpha = 2$, respectively, when $\alpha \rightarrow 2$.

\end{document}